\documentstyle{npbproc}
\def\beq{\begin{equation}}
\def\eeq{\end{equation}}
\def\beqr{\begin{eqnarray}}
\def\eeqr{\end{eqnarray}}
\def\r{{(r)}}
\def\1{{(1)}}
\def\2{{(2)}}
\def\Po{Poincar\'e }
\def\CS{Chern-Simons }
\def\eps{\epsilon}
\def\p{\varphi}

\begin{document}
\title{CLASSICAL DYNAMICS OF POINT PARTICLES IN 2+1 GRAVITY}
\author{Andrea Cappelli}
\address{Theory Division, CERN, Geneva, Switzerland
  (On leave from INFN, Firenze, Italy)}

\runtitle{Classical Dynamics in 2 + 1 gravity}
\runauthor{A.Cappelli}
\volume{99A}
\firstpage{1}
\lastpage{1}

\begin{abstract}
The relation between Einstein gravity
and the \CS gauge theory of the \Po group
is discussed at the classical level.
The general form of the gauge field is described in the presence
of point sources, corresponding to non-abelian anyons.
They have arbitrary motion apart from particle exchanges, which are
governed by the braid group.
The gravitational classical interaction appears when a specific
gauge fixing is taken, such that the metric is smooth and invertible,
with proper asymptotic behaviour.
Once the asymptotic motion is fixed, the classical two-body scattering
problem is well defined.
\end{abstract}

\maketitle

\section{Introduction }

Three-dimensional gravity is an interesting laboratory for understanding
both classical and quantum properties of field theories with
repa\-rametrization invariance \cite{DJH,Witten}.
Simple examples and exact solutions allow a non-verbose approach to
their many puzzling issues.
Besides, it has physical applications to cosmic strings in four
dimensions \cite{CTC}.

Here we shall limit ourselves to introducing matter by $N$ (dynamical)
point sources,
\beq
S=-{1\over 16\pi G_N}\int d^3 x \sqrt{g} {\cal R} -
\sum_{\r=1}^N m_\r \int ds_\r
\label{hilbert}\eeq

In the next section, we recall that this theory can be rewritten
as a \CS gauge theory of the \Po group $ISO(2,1)$, by using
the first-order (Palatini) formalism \cite{Witten}.
Namely, the long-standing hope of relating gravity to gauge theory
becomes a reality.
The peculiar aspects of this relation form the body of the following
discussion\footnote{
Based on the work done in collaboration with M. Ciafaloni
and P.Valtancoli \cite{CCV,Spin}.}.

At the quantum level, the \CS approach suggests that $2+1$ gravity is
renormalizable, in spite of non-renormalizability by power counting
of the Einstein-Hilbert action (\ref{hilbert}).
Actually, the two approaches are related by a non-linear change of
variables, involving a dimensionful scale, thus the two different
perturbative expansions correspond to different ``phases'' of gravity,
with unbroken $\langle g_{\mu\nu} \rangle =0$ and broken
$\langle g_{\mu\nu} \rangle \sim \eta_{\mu\nu}$
reparametrization invariance, respectively.

We would like to show that an insight in this still vague issue
comes from the classical problem of determining the motion of
point particles under their own field.
Actually, this can be tackled and its solution
shows some characteristic differences between gauge theory and gravity.

The problem of motion in gravity was first studied by
Einstein,Infeld and Hoffmann \cite{EIH} in four dimensions,
and it has never been solved beyond perturbation
theory, due to its formidable non-linearity.
In three dimensions, some exact many-particle space-times
were found by Deser, Jackiw and `t Hooft \cite{DJH},
owing to the following simplifing features:
i) there is no on-shell graviton,
ii) nor Newton static potential,
and iii) the space is flat outside sources.
Still, exact solutions were missing for two or more moving
particles. This is recalled in sect. 3 and fully discussed
in the reviews \cite{Jackiw}.

On the other hand, the anyon problem in \CS theory can be easily
solved completely \cite{CCV}. Amazingly, the motion is almost
arbitrary, because the trajectories of particles are gauge degrees
of freedom and can be deformed at will.
This solution is shown in sect. 4, in a generic gauge.
The unique constraint is that topologically inequivalent trajectories
({\sl i.e.} particle exchanges) are related by the braid group.
This acts on the topological observables, which are the holonomies of
the gauge field around each particle \cite{Carlip,Gerbert}.

In sect. 5, we show that the anyon problem indeed corresponds to
gravity by choosing an appropriate gauge-fixing, in which
the special properties of the
metric field with respect to the gauge field can be met\footnote{Somewhat
similar statements are also found in the refs.\cite{LWitten,Grignani}.}.
Namely, the metric should be smooth and invertible,
outside particle trajectories, and asymptotically constant.
Instead, the simpler anyon solutions give, in general,
non-invertible metrics, which, by definition, are not metrics.
The (singular) gauge transformation between the two cases
was called $\Lambda$-transformation \cite{CCV}.

Therefore, the main difference between gauge theory and gravity lies in the
relevance of the gauge choice, a problem complementary or ``dual''
to the solution of the equations of motion.
Actually, it is the hardest one in our approach, because it
brings back non-linearities into the problem.

Of course, the complete details of the trajectories ({\sl i.e.} of
the gauge) should not matter for the quantum theory, which
should be locally gauge invariant.
However, the asymptotic gauge-fixing is somehow unavoidable,
because it allows to identify the ``asymptotic states''
(asymptotic metric), and to define the two-body scattering.
The holonomies for open paths going to $\pm$ infinity
are similarly well defined, and yield new observables.
The time-like open path gives the scattering angle, which we
obtain exactly for massless particles.

In the conclusion, some open problems suggested by this classical
analysis are discussed.

\bigskip
\bigskip

\section{Gravity in the First-Order Formalism and \CS Theory}

In the usual second-order formalism $(g,\Gamma[g])$, the metric
is the field variable. In the first-order formalism
$(e,\omega)$, two field variables are introduced, the basis vectors
of the moving frame, or dreibeins $e^a_{\ \mu}$,
and the ``spin'' connection $\omega^a_{\mu b}$ for the associated
local Lorentz invariance.
They are related by the equations
\beqr
g_{\mu\nu} &= e^a_{\ \mu}\eta_{ab} e^b_{\ \nu} \qquad \quad
{\rm (soldering \ \ condition)} \label{gmunu}\\
\Gamma^\lambda_{\mu\nu} &=(e^{-1})^\lambda_{\ a}
\left( \delta^a_b\partial_\mu +\omega^a_{\mu b} \right) e^b_{\ \nu} \quad
{\rm (metricity)}
\label{Gconn} \eeqr
The first equation states how to connect the Lorentz manifold to
space-time, the second one defines the parallel transport which
preserves the metric, $D_\mu [\Gamma]\ g_{\alpha\beta} =0$.
A third equation involves the particle trajectory $\xi(\tau)$.
The space-time momentum
\beq
p^\mu_\r \equiv m_\r \dot\xi^{\mu}_{(r)} \ \ ,\ \ \ \ \ \ \ \ \
\left(\dot\xi\equiv{dx\over d\tau} \right)
\label{pmu}\eeq
is related to the Lorentz momentum $P^a$ by
\beq
p^\mu_\r \ e^a_\mu (\xi_\r) \ = \ P^a_\r \ \ ,({\rm particle} \ \
{\rm kinematics})
\label{Pa}\eeq

Thanks to the three-dimensional identity
\beq
\eps^{\mu\nu\rho}\ \eps_{abc}\ e^a_{\ \mu} e^b_{\ \nu} e^c_{\ \rho}
=\vert e\vert \qquad \qquad (D=3)
\label{determ}\eeq
the Einstein action (\ref{hilbert}) can be rewritten\footnote{
We consider here spinless particles, $J^a_\r =0$.}
(hereafter \hbox{ $8\pi G_N\equiv 1$})
\begin{eqnarray}  S=
&-& {1\over 2}\int d^3 x \ \epsilon^{\mu\nu\rho} \ \epsilon_{abc}
   e^a_{\ \rho} \left( \partial_{[\mu} \omega_{\nu]} +
   \omega_{[\mu}\omega_{\nu]} \right)^{bc} \nonumber\\
&-& 2 \sum_{(r)} \int d\tau  \left[ \dot\xi^\mu
   \left( P_a e^a_{\ \mu} -
         {1\over 2} J_a \eps^{abc}\omega_{\mu bc} \right)\right]_{(r)}
   \nonumber\\
\ \ =& \int &\left\langle A, \left({\rm d}A +{2\over 3} A^2\right)
   \right\rangle - 2\sum_\r \oint_\r A \nonumber\\
\ &\ & \label{CSaction}\end{eqnarray}
The explicit \CS form in the last equation follows by introducing a
gauge connection taking values in the \Po group $ISO(2,1)$
\beq
A_\mu=e^a_{\ \mu} {\cal P}_a \ -
{1\over 2}\eps^{abc}\omega_{\mu bc} {\cal J}_a \ \ ,
\label{A}\eeq
whose generators satisfy
\beqr
&[{\cal J}_a,{\cal J}_b]&=\epsilon_{abc}{\cal J}^c ,\ \
[{\cal J}_a,{\cal P}_b]=\epsilon_{abc}{\cal P}^c ,\nonumber\\
&[{\cal P}_a,{\cal P}_b]&=0
\eeqr
Furthermore, the gauge indices are contracted by using
the invariant non-degenerate metric
\beq \langle {\cal J}_a,{\cal P}_b \rangle = \eta_{ab} ,\ \
\langle {\cal P}_a,{\cal P}_b \rangle = 0 ,\ \
\langle {\cal J}_a,{\cal J}_b \rangle = 0 \
\label{trace}\eeq
Note that in the \CS theory $e^a_{\ \mu} \to e^a_{\ \mu}/G_N$,
thus $dim(e)=dim(\omega)=1$, and the dimensionless gauge
coupling constant is not quantized because \hbox{$\pi_3(ISO(2,1)) =0\ $}
\cite{Witten}.

The transformation (\ref{hilbert}-\ref{CSaction}) had
miraculous results.
We started from the highly non-linear theory (\ref{hilbert}),
not renormalizable by power counting
\hbox{$( dim G_N = -1)$},
and we arrived at the polynomial action (\ref{CSaction}),
which is renormalizable by expanding around the classical solution
\beq
e^a_\mu \sim 0 \ \ ,\ \ \omega^a_{\mu b} \sim 0
\label{unbroken}\eeq
which is reparametrization invariant.

Let us pause and think what this ``phase'' of gravity with
unbroken symmetry might be\footnote{
Following Witten \cite{Witten,TFT}.}.

Let us consider a quantum theory of gravity where the metric
is the fundamental field $\hat{g}_{\mu\nu}$. Next, let us ask what is
$\langle \hat{g}_{\mu\nu}\rangle$, the vacuum expectation value,
or better said, the semi-classical mean value.

The {\it broken phase}
$\ \langle \hat{g}_{\mu\nu}\rangle \sim \eta_{\mu\nu} \ $
(at least for large distances)
is the one of the classical limit, with geometry and light-cones.
This is better described by the Einstein action (\ref{hilbert}),
which however has the problem of non-renormalizable fluctuations
$\hat{g}_{\mu\nu} = \eta_{\mu\nu} + \hat{h}_{\mu\nu},
\hat{h}_{\mu\nu} \ll 1$, with reparametrization invariance broken to
Poincar\'e global symmetry.

The {\it unbroken phase} is instead characterized by
$\langle \hat{g}_{\mu\nu}\rangle \equiv g_{\mu\nu}=0$.
This equation is however not completely right, because
$g^{-1}_{\mu\nu}$ does not exist,
thus $g_{\mu\nu}$ cannot be called a metric and geometry is not
well-defined. Thus the metric cannot be a fundamental
field in this phase, or the other way around, this phase cannot be
seen by theories built on the metric field.
We have in mind a vague analogy with gauge theories like
QCD, descibed in terms of pions in the sigma-model or by
quarks and gluons, respectively.

The unbroken phase of $2+1$ gravity is naturally described by
the \CS theory \cite{Witten}.
This has a good ultra-violet behaviour, and the fundamental
gauge fields $(e,\omega)$ fluctuate around vanishing v.e.vs..
The metric is not present in eq.(\ref{CSaction}), thus the theory is
topological invariant.

However, in order to interpret this as a theory of gravity,
one has to go backward in the steps eqs.(\ref{gmunu}-\ref{determ})
and {\it define} the metric by the soldering condition (\ref{gmunu}).
This implies additional constraints on a naive treatment of the
\CS theory. The condition (\ref{gmunu}) breaks topological
invariance to reparametrization invariance, and it requires a
classical (long-distance) limit of the theory dominated by smooth gauge
configurations, yielding invertible metrics
\beq
\langle \hat{e}^a_{\ \mu} \rangle \sim\ {1 \over G_N} \delta^a_\mu
\label{vev}\eeq
where ${\hbar G_N}$ is the minimal length for the breaking
to take place.
Notice that a dimensionful scale is not present in the
(non observable) pure gauge theory (or pure gravity), but it is
introduced by the mass of the particles, owing to eq.(\ref{pmu}).

We believe that this is the main question, how the ``good''
\CS theory manages to reproduce geometry and the special role
played by the Minkowski metric. Actually, in classical gravity
the Minkowskian asymptotic behaviour is an additional physical condition
{\it not} implied by the local Einstein equations.
The natural guess is that this metric should emerge from the solution of
the quantum theory.
\bigskip

As a side remark, let us show that the first-order approach
does not lead to a renormalizable theory in four dimensions,
thus additional theoretical inputs are needed.
We repeat the previous transformation on the Einstein action
(\ref{hilbert}), using now the identity
\beq
\eps^{\mu\nu\rho\sigma}\ \eps_{abcd}\ e^a_{\ \mu} e^b_{\ \nu}
e^c_{\ \rho} e^d_{\ \sigma} = \vert e\vert \qquad(D=4)
\eeq
and we obtain
\begin{eqnarray}  S &=& \nonumber\\
&-& {1\over 8}\int d^4 x \epsilon^{\mu\nu\rho\sigma}
    \epsilon_{abcd}    e^a_{\ \mu} e^b_{\ \nu}
    \left( \partial_{[\rho} \omega_{\sigma]} +
   \omega_{[\rho}\omega_{\sigma]} \right)^{cd} \nonumber\\
&-& 2 \sum_{(r)} \int d\tau
   \left( \dot\xi^\mu P_a e^a_{\ \mu} \right)_\r
\label{fourdim}\eeqr
The action is again polynomial, but it has no term quadratic in the
fields. Therefore we cannot prove perturbative renormalizability,
because we cannot expand around $e\sim\omega\sim 0$.
The unbroken phase of gravity cannot be seen in the basic field variables,
and one has to invoke more hypothetical models in order
to keep this picture, like Topological Field Theories
\cite{TFT} and Pregeometric Models \cite{AV}.

Another possibility is to set up a non-pertur\-bative quantization method
for the action (\ref{fourdim}).
This is Ashtekar's approach \cite{Ashtekar}.
The proposal which attracted more attention is String Theory,
which indeed reproduces gravity in the classical {\it and}
high-energy limit \cite{ACV}.

Such additional theoretical structures are not needed in
three-dimensional gravity, which has the virtue of simplicity.
Physical problems common to four dimensions can be addressed directly,
and hopefully solved.

\bigskip
\bigskip

\section{The Einstein-Infeld-Hoffmann Problem}

Let us recall the problem of motion for point-like matter sources
in gravity.
The field equations of motion follow by varying the action
(\ref{hilbert}) w.r.t. the metric
\beqr
G_{\mu\nu}  = {\cal R}_{\mu\nu} - \frac{1}{2} {\cal R} g_{\mu\nu}
= T_{\mu\nu}, \nonumber\\
T_{\mu\nu} = \sum_\r { p^{(r)}_{\mu} p^{(r)}_{\nu} \over m_\r }
{ \delta^{(2)} ( \vec{x} - \vec{\xi}^{(r)} (t) ) \over \sqrt{g}}
{d\tau\over dt}
\label{einstein}\end{eqnarray}
where the trajectories are parametrized by
$\xi^\mu =(t,\vec{\xi}(t))$, and the momenta are given in
eq.(\ref{pmu}).

\begin{figure*}[t]
  \vspace{8cm}
  \caption{(a) Minkowski space with excised region, and (b) its
tail representation in the \CS solution. The trajectory of a test
particle is also indicated.}
\end{figure*}

In gauge theories like electrodynamics, the motion $\xi^\mu$ is
subjected to the independent particle equation (the Lorentz force).
In gravity instead, the particle (geodesic) equation
is not independent, but follows from
repara\-me\-trization invariance of the field equations
\beqr
0= D_{\mu} T^{\mu\nu} \nonumber\\
= \sum_{(r)}
\left[ {dp^\nu \over d\tau} + \Gamma^\nu_{\alpha\beta}[\xi]
p^\alpha p^\beta \right]_\r
{\delta^{(2)} ( \vec{x} - \vec{\xi}_{(r)}) \over m_\r\sqrt{g}}
{d\tau\over dt} \nonumber\\
\label{conserv}\eeqr

The difference is that in electrodynamics the particle equations are
gauge invariant, while in gravity the geodesic equation (\ref{conserv})
is repara\-me\-tri\-zation dependent\footnote{
See e.g. chapters 65 and 106 of ref.\cite{Landau}.}.

The Einstein equations (\ref{einstein}) give the field as a functional
of the particle motion $\{ \xi_\r \}$ and eq.(\ref{conserv}) gives
the feedback, resulting in a strong non-linear problem.
Moreover, the factor $\sqrt{g}$ implies that we cannot even write
down the equations without a guess of the solution.
This explains why this problem has only been solved perturbatively
or for a single source.

In three dimensions it simplifies considerably, because the Ricci
tensor determines uniquely the Riemann curvature
\beq
R_{\mu\nu,\rho\sigma} = - g \eps_{\mu\nu\lambda}\ \eps_{\rho\sigma\gamma}
\ G^{\lambda\gamma}\ \ \ , \quad\qquad (D=3)
\label{riemann}\eeq
Therefore, the classical space-times are flat apart from
singularities at the sources. For example \cite{DJH},
a static particle of mass
$m$ produces the following metric, in coordinates $x^\mu\equiv(t,r,\p)$,
\beq
ds^2 = dt^2 - dr^2 -r^2 \alpha^2 d \p^2 \ \ ,\quad 0\leq \p < 2\pi\ ,
\label{cone}\eeq
This is a cone because the circumference-to-radius ratio
$2\pi\alpha\equiv2\pi [ 1-( m/2\pi)] <2\pi$,
{\sl i.e.} the deficit angle is equal to the mass.
Alternatively, it can be represented as Minkowski space with an
{\it excised region}
\beq
ds^2 = dT^2 - dR^2 -R^2 d \phi^2 \ \ ,\quad 0\leq \phi < 2\pi\alpha ,
\label{cutcone}\eeq
where $\phi=\alpha\p,T=t,R=r$.
The $X$-coordinates are singular,
because the two edges of the excised region are indentified by a
finite rotation (see Fig.1a)
\beq
X^\mu_+ =\left(e^{m{\cal J}_0} \right)^\mu_{\ \nu} X^\nu_-
\label{rotation}\eeq
which corresponds to a delta term in the metric, not written in
eq.(\ref{cutcone}). Equation (\ref{rotation}) is the well-known
{\it matching condition} of Deser, Jackiw and 't Hooft
\cite{DJH}\footnote{
Quantum mechanics and scattering on this space were discussed
in refs. \cite{Quantum}.}.

The metric of a particle with momentum
$p$ can be obtained by performing a Lorentz boost $B(v)$.
It is again Minkowskian, with generalized
matching condition given by the \Po transformation
\beqr
X_+ -\xi(t) = e^{\ p\cdot{\cal J}} \left( X_- - \xi(t) \right)
\ \ ,\nonumber\\
\left(e^{p\cdot{\cal J}}  = B(v) e^{m{\cal J}_0} B(v)^{-1}\right)
\label{mc}\eeqr

For many particles, the flat space outside the sources can be taken
again Minkowskian, with an excised region for each particle --
neglecting for the moment their possible overlappings.
This picture was put forward by the previous authors as
a general solution of the motion.
While simple and appealing, this solution is not acceptable
because the metric is singular at the excised regions.
Moreover, it leads to a puzzling free motion for two
particles, because excised regions can never overlap.

\bigskip
\bigskip

\section{The Many-Anyon solution in \CS Theory}

As we said, singular metrics cannot be handled in the Einstein
theory, but are natural in the gauge theory of $(e,\omega)$.
The previous Deser-Jackiw-'t Hooft (DJH) picture of
multi-excised Minkowski space better applies to the following
solution of the anyon problem, in the simplest gauge.
More general solutions with singular metric will be found
in other gauges, which allow for {\it arbitrary} motion of the particles.
We shall also see what happens when two excised regions meet.
In sect. 5, we shall discuss how to close the excised regions and obtain
a smooth metric, namely how to undo the transformation
from eq.(\ref{cone}) to eq.(\ref{cutcone}) in the many-particle
case.

\bigskip\bigskip
{\bf 4.1. Equations of Motion}
\bigskip

The \CS field equations are read off from the action (\ref{CSaction})
\begin{eqnarray}
R_{\mu\nu\ b}^{\ \ a}&\equiv&
    ( \partial_{[\mu} \omega_{\nu]}  +
    [ \omega_{\mu}, \omega_{\nu} ] )^a_{\ b} \nonumber\\
& =& - \epsilon_{\mu\nu\lambda} \epsilon^a_{\ bc}
\sum_r v^\lambda_r P^c_\r \delta^2 ( x - \xi_\r) \ ,
\label{rie}\\
T_{\mu\nu}^{\ \ a}&\equiv &
\partial_{[\mu} e_{\nu]} + \omega_{[\mu}e_{\nu]} = 0
\label{torsion}\eeqr

These are the Cartan structure equations of the first-order formalism.
The first one expresses the Riemann tensor in terms of the sources,
as eq.(\ref{riemann}), while eq.(\ref{torsion}) gives a
vanishing torsion\footnote{
For spinning particles, the torsion is proportional to the spin sources,
thus the \CS theory corresponds to the more general Einstein-Cartan
theory of gravity \cite{Spin}.}.

By introducing the $4\times 4$ representation
of the \Po group
\beq A_\mu =\left(
\begin{array}{c c} \omega^a_{\mu b} & e^a_\mu \\
                                  0 & 0 \end{array} \right),
\ \ {\cal J}_a =\left(
\begin{array}{c c} -(\eps_a)^b_{\ c} & 0 \\
                                   0 & 0 \end{array} \right)
\nonumber\eeq
one can easily see that the r.h.s. of eqs.(\ref{rie},\ref{torsion})
are the components of the (flat) connection $F_{\mu\nu}$
in the \CS theory.

First-order gravity and \Po gauge theory have the same
gauge transformations, provided the equations of motion are
used \cite{Witten}.
Therefore, the geodesic particle equation
should follow similarly from local \Po invariance of the \CS
field equations. Actually, this implies the identity
\beq
\epsilon^{\mu\nu\rho} {\cal D}_\rho F_{\mu\nu}=0, \ \ \ \ \
{\cal D}_\rho \ \ \equiv \partial_\rho \ \ +[A_\rho, \ \ ]
\label{bianchi}\eeq
which reads in components
\beqr
\dot{P}^a_{(r)} + \dot\xi^\mu_{(r)} \omega^a_{\mu b}P^b_{(r)}=0
\label{geoeq}\\
\epsilon^a_{\ bc} P^b_{(r)} e^c_{\ \mu} p^\mu_\r =0
\label{parteq}\eeqr
The first component is recognized as the geodesic equation
(\ref{conserv}) in $(e,\omega)$ variables, by using eqs.(\ref{Gconn},
\ref{Pa}).
The second one is similar to the kinematical relation (\ref{Pa}),
but less strong because it does not fix the mass of the particles.
This breaks scale invariance and cannot follow from the \CS theory,
it should be added as an additional requirement, like the definition
of the metric discussed before.

Notice that eqs. (\ref{geoeq},\ref{parteq}) are ``charge''
conservation equations in \CS, where the non-abelian charge is
given by the Casimir $P^2=m^2$ (and $J\cdot P =m\sigma$
for non zero spins $\sigma_\r$). They become dynamical equations
due to the identification of gauge and space-time manifolds
necessary for gravity.
Therefore, we should not include the (non-abelian) Lorentz equation
for the particles, as suggested by the analogy with
the Yang-Mills theory, and done for anyons in
condensed matter physics \cite{Bala}.

\bigskip\bigskip
{\bf 4.2. The Solution}
\bigskip

To summarize, the \CS equations of motion are
(\ref{Pa}), (\ref{rie}), (\ref{torsion}) and (\ref{geoeq}).
The general solution will be
characterized by constant Lorentz momenta
\hbox{ $P^a_{(r)} \ = \ m_{(r)} U^a_{(r)} \ = \ $ } const,
and parallel (non-covariant) velocities
$V^i_\r =U^i_\r / U^0_\r  (i = 1,2)$ .
The generalization to non-parallel ones and non-constant
$P^a$\  's will be discussed later.
The solution exists for arbitrary trajectories
$\{ x^i \ = \ \xi^i_{(r)} (t) \}$,
which can be written
\beq\begin{array}{l l l}
X^2 (\xi^{\mu}_{(r)}) & = & B_{(r)} = {\rm const.}\\
X^1 (\xi^{\mu}_{(r)}) & = & V_{(r)} X^0 (\xi^{\mu}_{(r)})
\end{array}\label{traj}\eeq
where $X^a (x^\mu)$ are arbitrary functions
of the coordinates $x^\mu$, invertible at fixed time,
\beq
J_\r \ = \left\vert {\partial (X^1 \ - \ V_{(r)} X^0, X^2) \over
                  \partial ( x^1, x^2)} \right\vert_\r > 0
\label{jacob}\eeq

Let us present the solution in the simplest gauge
$X^a =\delta^a_\mu x^\mu$,
\begin{eqnarray}
\omega_{\mu} & = & \sum_\r \omega_{\mu}^{(r)} \label{omegasol}\\
e^a_{\ \mu}  & = & \partial_{\mu} X^a +
\sum_\r \omega^a_{\r\mu b} ( X - B_{(r)} )^b
\label{esol}\end{eqnarray}
where, for each particle $\r$,
\beqr
\omega_{\r\mu} &=&(P_\r\cdot{\cal J})
\left( \partial_{\mu} \Theta ( X^2 -  B_{(r)} )\right) \nonumber\\
              &\times&\Theta(V_{(r)} X^0 - X^1)
\label{omegar}\eeqr
where $\Theta$ is the step-function.
This solution is linear in the sources and satisfies the superposition
principle.
Any particle has associated a branch cut on the left, or ``tail'' (Fig.1b),
the support of the spin connection.
The tail location is a gauge choice and it can be freely rotated,
provided another tail is not met.

Notice that $\ \ (1 + \omega_{(r)})\ \ $
is an infinitesimal rotation around the constant vector $P_{(r)}$,
therefore the term $\omega_\mu P$ in the geodesic equation
(\ref{geoeq}) drops out, and our ansatz $P^a_{(r)}=$const.
is consistent.
The field equations are easily verified, because the quadratic
terms vanish for separated tails.

The solution for generic trajectories can be obtained by
performing the arbitrary coordinate transformation
$X^a(x^\mu)$.
Actually, the equations of motion and the solution have an
intrinsic expression in terms of differential forms, so that
their relations are reparametrization invariant.

In the general case, the $P^a$'s stay
constant but the $p^\mu$ are not, thus the trajectories in space-time
bend and particles do interact.
We clearly see that dynamics in three dimensions is not determined
by the local equations of motion, but by the choice of gauge,
subjected to more subtle global metric conditions.

Before coming to this point, let us show that our solution in
the simplest gauge
$X^a=\delta^a_\mu x^\mu$ reproduces Deser-Jackiw-`t Hooft
excised space-time.
The geometry of such singular metric can be understood
by studying the geodesics, {\sl i.e.} by sending a beam of
test particles.
Their geodesic curves $x^\mu (\tau)$ also satisfy eq.(\ref{geoeq}),
which can be integrated once
\beq e^a_{\ \nu} {\dot{x}}^{\nu} ( \tau ) =
P \left( \exp - \int^{\tau}_{\tau_0} dx^{\mu} \omega_{\mu} \right)^a_{\ b}
\ U^b \label{geod}\eeq
where $P$ is the path -ordering along the geodesic, and $U$ is a constant
velocity. In the region outside tails $\omega_\mu=0$, thus the geodesics are
straight lines. Across the tail, the tangent vector makes a jump.
Near the r-th tail we can substitute eq.(\ref{esol}) in eq.(\ref{geod}),
integrate again and obtain the matching condition (\ref{mc}),
for a point just above $(X_+)$ and just below $(X_-)$
the tail (fig.1) \cite{CCV}.

\bigskip\bigskip
{\bf 4.3. Constants of Motion from the}

{\bf Holonomies}
\bigskip

The observables of the $ISO(2,1)$ Chern- Simons theory are the loop integrals
\beq U_\Gamma(y,x)=P \exp \left(-\int_\Gamma
\omega\cdot{\cal J} + e\cdot {\cal P}\right)
\label{ugamma}\eeq
where $\Gamma$ is an oriented path from $x$ to $y$.
Gauge invariant quantities can be obtained from closed loops, by looking
for quantities invariant under $U \to gUg^{-1}$.
By parametrizing
\begin{equation} U_\Gamma(x,x)=
\left( \begin{array}{c c}  L=e^{-w\cdot{\cal J}} & q\\
                                  0 & 1 \end{array} \right)
\equiv \left( L, q, 1 \right)
\end{equation}
we find the following two invariants
\begin{eqnarray}
\sqrt{w^2} & \ \leftrightarrow \ \ &
{\rm Tr}_{(1)} L = 1+2\cos(\sqrt{w^2}) \label{angle}\\
\sigma= & {\displaystyle{q^a w_a\over \sqrt{w^2}}} & \ \
\label{spinproj}\end{eqnarray}
$(\sigma=q^a w_a {\rm \ \ for \ \ } w^2=0)$.
The Lorentz invariant is the angle of the
(pseudo)-rotation $\sqrt{w^2}$,
the spin invariant is the projection of the translation
$q^a$ on the rotation axis \cite{Gerbert}.

The loops are also invariant under smooth deformations of the metric,
because they are given by the integral of a one-form.
The particle motion being a kind of topological
deformation, it will not affect their value. Thus the
invariants (\ref{angle},\ref{spinproj}) are
constants of motion of the particle plus field dynamics
\cite{Moncrief}.

Any particle $\r$ has associated an elementary holonomy $U_\r (x_\r, x_\r)$,
for a tiny loop surrounding it counterclockwise, with basepoint $x_\r$
near the particle, off the tail,
\begin{eqnarray}
U_\r (x_\r, x_\r) &=&\exp(-\int_{\Sigma_\r}d\sigma^{\mu\nu} F_{\mu\nu})
\nonumber\\
&=&\left(L_\r\equiv e^{-{\cal J}\cdot P_\r}, J_\r \ , \ \  1 \right)
\label{ur}\end{eqnarray}
which can be computed from the field equations \cite{CCV,Spin}.
Its invariants
(\ref{angle},\ref{spinproj}) are just the Casimirs of the particle
$ w^2=P^2_\r= m^2_\r$, and $\sigma=(P\cdot J/m)_\r= \sigma_\r$,
independent of time as announced (hereafter $\sigma_r$=0).

Let us consider now two particles and the holonomy for
the loop encircling them once counterclockwise.
{}From the solution (\ref{omegasol},\ref{esol}) we can compute its
corresponding invariants, which are
the rotation ${\cal M}$ and the spin $S$ invariants \cite{CCV},
\beqr
\cos{{\cal M}\over 2}
= c_\1 c_\2 - s_\1 s_\2 {P_\1 \cdot P_\2\over m_\1 m_\2}\\
S \ \sin{{\cal M}\over 2} = 2 s_\1 s_\2 \epsilon_{abc} (B_\1 -B_\2)^a
{P_\1^b P_\2^c \over m_\1 m_\2}
\nonumber\label{oloop}\eeqr
where $c_{(i)}= \cos{m_{(i)}\over 2}, s_{(i)}=\sin{m_{(i)}\over 2}$.

These two formulas were obtained by Deser, Jackiw and
't Hooft \cite{DJH}. Their recipe of composing
matching conditions for each particle finds
the correct meaning in the \Po holo\-no\-mies\footnote{
See also ref.\cite{Carlip}.}.
${\cal M}$ and $S$ are two of the infinite
conservation laws of the Chern-Simons theory.

In non-invariant terms, Deser, Jackiw and 't Hooft argued
that there is a choice of coordinates,
which at large space-like distances
\hbox{$\vert\underline{x}\vert \gg \vert\underline{B}\vert$}
is Minkowskian with a single excised region and jump in time determined by
${\cal M}$ and $S$ in eq.(\ref{oloop}), a kind of center-of-mass system.
Actually, for $G_N m_\r \ll 1$,
${\cal M}$ and $S$ reduce to the usual formulas of special relativity
for the total invariant mass and the total angular momentum.
Higher corrections were interpreted as global contributions
of the gravitational field.
However, this reasoning on singular metrics encounters a new
difficulty, because inertial or accelerating center-of-mass systems
should be distinguished by the asymptotic behaviour of the metric,
which is undecidable in this case.

\bigskip\bigskip
{\bf 4.4. Particle exchanges and the Yang-Baxter equation}
\bigskip

Let us now discuss how the solution extends to the case of
non-parallel velocities, leading to crossings of tails.
Without loss of generality, consider the case of particle (2) crossing
the tail of particle (1) (fig.2).
$P^a_{(2)}$ is no longer constant, and it evolves according to the integrated
eq.(\ref{geoeq})
\beq
P_{(2)} ( x^{+}) = P \exp\left(
-\int^{x^+}_{x^-}\omega_\1\right) P_{(2)} ( x^{-} )
\eeq
where $x^+\ (x^-)$ correspond to $t>0 \ (t<0)$, the collision being at $t=0$.
If we ignore the feedback of (2) on (1) (test-particle limit), this equation
gives the matching condition of the geodesic eq.(\ref{mc})
\begin{eqnarray}
P_{(1)} ( x^+ ) & = & P_{(1)} ( x^- ) \nonumber\\
P_{(2)} ( x^+ ) & = &  L_{(1)} \ P_{(2)} ( x^- )
\label{jump}\end{eqnarray}
Actually this result is also valid when both particles are dynamical
\cite{CCV}.
It can be proven by using the integrated form of the field equations
given by the non-Abelian Stokes theorem \cite{Bala}.

\begin{figure}[t]
  \vspace{7cm}
  \caption{The particle exchange operation $\sigma_{12}$, for the
indicated orientation of tails.}
\end{figure}

One should not think of the crossing of tails as the scattering
of two particles. It might look so in this
particular gauge $X^a=\delta^a_\mu x^\mu$, but
it is possible to put it in an invariant form,
which holds for all solutions in {\it arbitrary} gauge $X^a(x^\mu)$.
Thus it involves only the topology of the trajectories
and manifestly gives no conditions on the $p^\mu_\r$'s.
Therefore, a correct name for the process of tail crossing
is {\it particle exchange}.

\begin{figure*}[t]
  \vspace{7cm}
  \caption{Sequence of exchanges for the Yang-Baxter equation.}
\end{figure*}

More precisely, let us define a particle exchange operator
$\sigma_{12}$ (fig.2) acting on the tensor
space $V_{(1)} \otimes V_{(2)}$, which contains the elementary
Poincar\'e holonomies of the two particles.
They should have a common base-point,
thus $U_\2$ in eq.(\ref{ur}) is translated to $B_\1$,
\beqr
U_\1&=&\left( L_\1, \ 0,\ 1\right), \nonumber\\
U_\2&=&\left(L_\2, \ (L_\2 -1)(B_\1 - B_\2), \ 1\right)
\eeqr
The effect of particle $\2$ crossing the tail of $\1$
can be written in the form
\beq
\sigma_{12}\ :\ \ \left\{ \begin{array}{l}
      U_\1 \rightarrow  U_\1 \\
      U_\2 \rightarrow  U_\1  U_\2  U_\1^{-1}
\end{array} \right.
\label{sigma}\eeq
which is manifestly topological invariant.
It is partially gauge dependent, because
a different tail orientation, i.e. a different choice of basepoint,
would have given the operator $\sigma_{21}$, thus exchanging
the (asymmetric) role of the two particles.
On the other hand, the monodromy of the particle $(2)$ around $(1)$,
{\sl i.e.} the double braiding $\sigma_{12}\sigma_{21}$,
is gauge invariant\footnote{
Similar investigations of the Braid group were done in the refs.
\cite{Carlip},\cite{Bala}.}.

Next, the exchanges of $N$ particles follow
by composition of the operators $\sigma_{i,i+1} , i = 1,...,N-1$,
which generate the braid group ${\cal B}_N$.
They act on tensor spaces $V_\r$ of particle holonomies, all with the same
basepoint.
The generators $\sigma_{i,i+1}$ satisfy the Yang-Baxter equation
\begin{eqnarray}
\sigma_{ij} \sigma_{ik} \sigma_{jk} = \sigma_{jk} \sigma_{ik} \sigma_{ij},
\ \ \ \ k=j+1=i+2 \nonumber\\
\sigma_{i,i+1} \sigma_{j,j+1} = \sigma_{j,j+1} \sigma_{i,i+1},
\ \ \vert i-j\vert \geq 2
\label{YB}\end{eqnarray}
where the first equation acts on $V_{(i)} \otimes V_{(j)} \otimes V_{(k)}$
and the second one on
$V_{(i)}\otimes V_{(i+1)} \otimes V_{(k)}\otimes V_{(k+1)}$.
This equation expresses associativity of particle exchanges, as required by
the deformability of trajectories (fig.3).

The Yang-Baxter equation is not trivial in our non-Abelian case,
and it provides a consistency check of our previous analysis on crossings
\cite{CCV}.
Clearly, it should be an identity at the classical level, because
conditions on the momenta would mean a classical
motion only for some initial conditions.

At the quantum level, one should find a representation of the
Braid Group in the Hilbert space, and show that the physical states
are in the trivial singlet representation. This strong condition has not
yet been solved in the non-Abelian case \cite{Adamo,Smatrix}.

\bigskip
\bigskip

\section{Smooth metrics and the $\Lambda$-transformation}

Let us now look at the gravitational
problem, for which the behaviour of the metric $ g = e^T \eta e\ $
is important. We want to avoid the $\delta$-function singularities present
in the general solution (\ref{omegasol}-\ref{omegar}),
{\sl i.e.} to close the excised regions discussed before.

\bigskip\bigskip
{\bf 5.1. One Particle}
\bigskip

In the case of a static particle, the angular rescaling
from (\ref{cone}) to (\ref{cutcone}) can be interpreted
as the transformation
\beq
X = \exp\left({m\over 2\pi} \p (x) {\cal J}_0 \right) x  =
\ \Lambda(x) x
\label{llambda}\eeq
The $\Lambda$ rotation
varies from $- m/2$ to $ m/2$ when $\p$ varies from $-\pi$
to $\pi$ and thus generates a solution to the matching condition for the
$X$-variable in eq.(\ref{rotation}).
As a consequence, $x$ is continuous, as it should be.
These continuous coordinates $x^\mu$ correspond to a choice of gauge
for $X(x)$ in the general solution (\ref{esol}),
which gives the dreibein
\begin{eqnarray}
e^a_{\ \mu} &=& [(\partial_{\mu} + \omega_{\mu} ) \Lambda x ]^a \ =
\nonumber\\
&=& \Lambda^a_{\ b} ( \delta^{b}_{\mu} + {m\over 2\pi}
n^{b} n_{\mu} ) ,
\eeqr
where $n_{\mu} = ( 0, \epsilon_{ij} {\hat{x}}^j ) $.
After some computations, the metric becomes the one in
eq.(\ref{cone}), as expected.

For the moving particle, the $\Lambda$-transformation is obtained
by applying a Lorentz boost to the static transformation
(\ref{llambda}) \cite{CCV}.
A particularly interesting limit is the massless case, with fixed energy E,
obtained by letting
\beq m \rightarrow 0 \ \  ( \gamma \rightarrow \infty ),
\ \ {\rm with} \ \ m \gamma \ = \ E \eeq
The result is the Aichelburg-Sexl metric \cite{AS}
for a massless particle in three dimensions,
\beq ds^2 \ = 2 dudv - (dy)^2 + \sqrt{2} E \delta(u) \vert y\vert (du)^2
\label{gAS}\eeq
where the light-cone variables are
\beq u \ = \ \frac{t-x}{\sqrt{2}} , \ \ \ \ v \ = \ \frac{t+x}{\sqrt{2}}
\nonumber\eeq

Notice that the Lorentz contraction has transformed the smooth
angular dependence in eq.(\ref{llambda}) into a step-function,
\beq
\Lambda \to \left\{ \begin{array}{l l}
e^{-p\cdot{\cal J}/2} & -\pi < \p < - \pi/ 2 \\
1                     & -\pi/2 <\p < \pi/ 2 \\
e^{\ p\cdot{\cal J}/2}  &  \pi/ 2 <\p < \pi
\end{array}\right.
\label{zeromass}\eeq
producing a metric (\ref{gAS}) with a (now physical) singularity
at the wave-front, or shock wave (fig.4).

\bigskip\bigskip
{\bf 5.2. Geodesic Scattering}
\bigskip

The $\Lambda$-mapping is particularly useful to describe the motion of
a test particle in the one-particle metric described before. In fact we
already know that, in $X$-coordinates, the geodesics are straight lines,
if they do not cross the tail.

If $\phi$ denotes the azimuthal $X$-coordinate in the rest frame,
eq.(\ref{cutcone}), we obtain by inverting eq.(\ref{llambda}),
\beq
x^a (\tau) \ = \Lambda^{-1} \left(\frac{\phi(\tau)}{\alpha}\right)^a_{\ b}
\ ( U^b \ \tau \ + \ B^b )
\label{xtraj}\eeq
It is thus easy to discuss the asymptotic motion and scattering of the test
particle. The asymptotic velocity becomes
\beqr u_\pm ( \infty)
&=&\Lambda^{-1} ( \pm \frac{\pi}{\alpha}) u_\pm ( -\infty) \nonumber\\
&=& e^{\pm p\cdot{\cal J}/2\alpha} \ u_\pm(-\infty)
\label{asympt}\eeqr
where the $+(-)$ sign holds according to whether the geodesic runs above
(below) the field particle.
Special cases of eq.(\ref{asympt}) are:

i) The geodesic scattering angle off a particle of mass $m$ at rest
\cite{DJH}
\beq \theta_0 \ = \frac{m}{2\alpha} \ = {m\over 2}
\left(1-{m\over 2\pi}\right)^{-1}.
\eeq

iii) The scattering angle for both massless particles
\beq \tan \frac{\theta}{2} \ = \ \frac{E}{2}
\label{thetaAS}\eeq
where $E$ is the energy of the field-particle,
in agreement with the derivation from the Aichel\-burg-Sexl
metric \cite{AS}.

The above results show that the gravitational problem for test particles is
characterized by quantities  which are not purely topological, and in our case
are given in terms of the Christoffel connection by
\beq P exp {( - \int^{\infty}_{-\infty} \ dx_{\mu} \ \Gamma^{\mu} )}_{\pm} \
= \ {[ L(P) ]}^{\pm \frac{1}{2\alpha}}
\label{pathgamma}\eeq
where the path runs along the geodesic above (below) the point source
of momentum $P$.

One may wonder to what extent this result is gauge dependent.

In principle, one may choose to change the $\Lambda$-transformation in the
particle rest frame, by some reparametrization of the azimuthal variable
$\p$. This, however, will make the scattering angle in general dependent
on the direction of the probe in an arbitrary way. Since this violates the
physical notion of isotropy of space for a spinless particle, we shall
explicitly exclude it in the following.

In other words, we require on physical grounds the $\Lambda$-transformation to
be given (at large distance, i.e., for $r\to\infty$) by
eqs.(\ref{llambda}), which define the isotropic
single-particle metric in our framework. Then the result (\ref{pathgamma}),
due to the properties
of parallel transport, will be invariant under any local reparametrization
which is asymptotically consistent with our choice.

To summarize, in the singular $X$-coordinates the geodesic scattering
is ambiguous, because of the
possibility that a branch-cut is crossed. On the other hand, in the smooth
$x$-coordinate, the geodesic scattering angle is uniquely determined in the
rest-frame (with the isotropic choice) and therefore also in any other
frame, giving rise to the more general ``S-matrix'' in eq.(\ref{pathgamma}).

\bigskip\bigskip
{\bf 5.3. Two-Particle  Scattering}
\bigskip

Let us now consider the problem of finding the gauge fixing
$ X^a = \overline{X}^a ( x^\mu )$ which gives
a smooth metric $g=e^T \eta e$ in the case
of two dynamical particles.
This is more difficult than the previous one-particle
transformation (\ref{llambda}),
and we do not yet have a complete solution.
Nevertheless, we shall state the gauge conditions which,
in our opinion, are sufficient to obtain  a non-ambiguous result for
the scattering process.
These conditions admit a solution in the first perturbative order in
$Gm_\r$. Under the assumption that an exact solution exists, we shall
give an ansatz for the exact scattering angle in the massless case\cite{CCV}.

These gauge conditions are as follows:
\bigskip

--- Smoothness conditions.

The gauge fixing is supposed to be of the form
\beq
\overline{X}^A \ = \ {({\cal{T}} (x))}^A_{ B} \ x^B, \ \ \ (A=0,1,2,3)
\label{gf}\eeq
where ${\cal{T}} (x)$ is a Poincar\'e transformation
which builds a solution of the matching conditions for each particle,
\beq
{\cal{T}} \simeq {\cal{T}}_{(r)} (x) S_{(r)} (x) , \ \ \ \ \
x^\mu \simeq \xi^\mu_{(r)} (\tau),
\label{facto}\eeq
where
\begin{eqnarray}
{\cal{T}}_{(r)} \ &=& \ e^{B_\r \cdot {\cal P}}
\Lambda_\r \left(\p_\r\right)
e^{- B_\r\cdot {\cal P}}\ \ \ \ \ ( r = 1, 2 )\nonumber\\
\Lambda_\r &=&
\exp\left({\p_\r (x)\over 2\pi} \ P_{(r)} \cdot {\cal J} \right)
\label{tra} \end{eqnarray}
is the (properly translated) single particle transformation discussed in the
previous section, and $S_{(r)}$ is instead regular at that point.
{}From the definition (\ref{tra}) it follows that, close to $\xi^\mu_{(r)}$,
${\cal{T}}$ performs the change of variables which closes the
excised region, as in eq.(\ref{llambda}).
In the case of scattering $P_\1 \neq P_\2$,
the non-commutativity of the $\Lambda_{(r)}$'s is the main
difficulty in finding the exact form of ${\cal T}$.
\bigskip

--- Asymptotic conditions.

Further conditions correspond to the definition of asymptotic states.
The commutator of $\Lambda_{(1)}$ and $\Lambda_{(2)}$
becomes vanishingly small when the particles are far apart, thus
we can impose the initial condition
\beq
{\cal{T}} (x) \rightarrow {\cal{T}}_{(1)} (x) {\cal{T}}_{(2)} (x) \simeq
{\cal{T}}_{(2)} (x) {\cal{T}}_{(1)} (x)
\eeq
for fixed spatial coordinates and large negative times. This ensures that
the class of metrics we are interested in is asymptotically consistent with
the gauge fixing for a single particle given in the previous section.

By inspection, the momenta of the incident particles are
the constants $P_{(1)}, P_{(2)}$ of our solution,
\beq
p^\mu_{(1)} ( -\infty) = {\delta}^\mu_a P^a_{(1)}, \ \ \ \
p^\mu_{(2)} ( -\infty) = {\delta}^\mu_a P^a_{(2)},
\eeq
Moreover, one can check that the parallel trans\-port from (1) and (2)
to the central region \hbox{$\vert x\vert\ll T$} is trivial.
Therefore, in this region we can define, as in special relativity,
the total momentum,
\beqr
p^\mu = p^\mu_{(1)} + p^\mu_{(2)}, \nonumber\\
p_{\mu} = g_{\mu\nu} (-\infty)  p^\nu = \eta_{\mu\nu} p^\nu\ ,
\eeqr
the invariant mass squared $ s = {( p_{(1)} + p_{(2)} )}^2 $,
and finally the center-of-mass frame by \hfill\break
$(p_\1 +p_\2)^i=0,\ \ i=1,2\ $.

An additional asymptotic condition is needed, at fixed time and
large $\vert x\vert$, in order to avoid rotating frames at infinity. This
requires that the Christoffel connection $\Gamma_\mu$ is as
small as dimensionally allowed,
\beq
\Gamma_\mu = O \left( \frac{1}{\vert\underline{x}\vert} \right)
\ \ \ \ \ (\vert\underline{x}\vert
\rightarrow \infty, T {\rm \ fixed }).
\label{Rinf}\eeq
In contrast, in a rotating frame $\Gamma_\mu$ would be of order
$(\Omega^2 \vert\underline{x}\vert)$.
\bigskip

These gauge fixing conditions admit a non-trivial solution
in perturbation theory, \hbox{$GE_\r\ll 1$}.
This is analogous to the ``fast'', weak coupling, approximation used in
four dimensions \cite{EIH}.
The Lorentz part $\Lambda$ of the transformation ${\cal T}$
in eq.(\ref{gf}) is approximately of product type
\beq
\Lambda \simeq \Lambda_{(1)} \Lambda_{(2)} \simeq \Lambda_{(2)} \Lambda_{(1)}
\eeq
because the commutator terms are of 2nd order in the $GE_\r$'s. This takes
care of both the polydromy and initial state conditions to this order.
Finally there are no rotations at infinity, because the angles $\p_\r$
are body-fixed with the straight-moving particles, thus
eq. (\ref{Rinf}) is verified.

\begin{figure*}[t]
  \vspace{8cm}
  \caption{(a) Initial and (b) final state pattern of the
$\Lambda$-transformation for the scattering of two massless particles.
(The tail and wave-front singularities of $\Lambda$ are indicated
by curly and straight lines). }
\end{figure*}

The first-order center-of-mass scattering angle turns out to be \cite{CCV}
\beq
\theta = \frac{\sqrt{s}}{2} ( 1 + O(G m_{(r)})), \ \ \ \ (8\pi G_N = 1)
\eeq
in agreement with the classical limit of string scattering \cite{ACV}.
\bigskip

The zero mass, or high energy limit \hbox{$m_\r \ll \sqrt{s}$},
has also some simplifying features.
The $\Lambda$-trans\-formation has the form of a shock wave,
eq.(\ref{zeromass}), with no effect on times earlier than the
arrival of the particle. Thus we can superpose two one-particle
Aichelburg-Sexl metrics with Minkowski space in-between, both
before and after the collision of wave-fronts at $T=0$.
This is shown in fig.4, where the $\Lambda$ is constant in each sector
and takes the values $L_r^{\pm 1/2} =\exp (\mp p_\r\cdot {\cal J})$,
with $p_\1=(E,E,0),\  p_\2 = (E,-E,0)$.

Since the $P_\r$'s are constant, and therefore the discontinuities across the
wavefronts and tails are fixed, the $t>0$ $\Lambda$-mapping is
still piecewise constant
in the regions bounded by wavefronts and tails (\ref{zeromass}),
and uniquely determined by
the polydromy requirements (\ref{tra}) up to the following
Lorentz transformation
\beq
R^{-1}(\theta) = \lim_{\epsilon\rightarrow 0} \left.
P exp \left( - \int^{t = +\epsilon}_{t = - \epsilon} \Gamma_\mu
dx^\mu \right) \right\vert_{\underline{x} = 0}
\label{Sclass}\eeq
which gives the classical S-matrix in this case.
Notice that the $X$-coordinates stay constant at $T=0$, because the
two tails do not cross.

$R^{-1}$ is determinated by the coordinate condition that there are
no rotations at infinity at $T=0$ in
the $x$-coordinates external to the wave-fronts \cite{CCV}.
We obtain the rotation angle
\beq
\tan\frac{\theta}{2} = \frac{E}{2} = \frac{\sqrt{s}}{4} \ \ \ \ \ \
( 8\pi G_N = 1 ).
\label{formula}\eeq
Notice that this angle is the same as the geodesic one,
eq.(\ref{thetaAS}), but the corresponding S-matrices
(\ref{pathgamma},\ref{Sclass}) are quite different, in particular
the latter one is in agreement with energy conservation.

\bigskip
\bigskip

\section{Conclusions and Outlook}

We have shown that the first-order approach to three-dimensional
gravity, rephrased in the \CS gauge theory, allows a larger class of solutions
of the classical field theory, because reparametrization invariance
need not be completely fixed.

This has clearly shown that the trajectories of particles are
gauge degrees of freedom. They are arbitrary, apart from
the topological constraint given by the Braid group\footnote{
And the condition of eq.(\ref{jacob}).}.
Therefore the classical motion in gravity
is completely determined by the gauge choice.

There are no natural gauge choices in the classical \CS theory.
Similarly, the unbroken phase seems to be the natural one
in the quantum theory.

On the other hand, we have enphasized that many aspects of the
classical motion in the Einstein theory are gauge dependent,
and that there are physical gauges, for which the metric is
smooth, isotropic, and asymptotically inertial.
For gravitational scattering,
it is sufficient to choose the physical gauge
asymptotically (asymptotic breaking),
because the scattering angle can be written as the holonomy for an
open time-like path, eqs.(\ref{pathgamma}) and (\ref{Sclass}),
which is invariant under localized reparametrizations.

Therefore, from the classical point of view, the unbroken
phase of gravity at short distance and the broken one at
large distance are not incompatible, provided one understands
the quantum mechanism for this asymptotic breaking.

This can only arise from the coupling of quantized matter to gravity,
because the pure gravity case does not show it \cite{Witten}.
The known actions of matter coupled to gravity are based on
the equivalence principle, and need an invertible metric field,
therefore they cannot be used in the unbroken phase.
An $ISO(2,1)$-invariant coupling of matter
to first-order gravity is not yet known.
This problem can possibly be by-passed by
quantizing the degrees of freedom of the point sources,
in the effective action obtained by our general classical solution.
In this kind of Coulomb gas approach, the main problem to
start with is the invariance of Hilbert space
under the Braid group, as mentioned before \cite{Adamo,Smatrix}.

\bigskip

{\bf Acknowledgements}
\bigskip

I would like to thank my collaborators in this work, Marcello
Ciafaloni and Paolo Valtancoli, and Daniele Amati for
interesting discussions.
I also thank the Organizers of this Workshop and the
Departament ECM, University of Barcelona, for their kind hospitality.
\def\NP{Nucl. Phys.\ }
\def\AP{Ann. Phys. (NY)\ }

\end{document}